\documentclass[runningheads]{llncs}
\usepackage{graphicx}
\usepackage{multirow, makecell}
\usepackage{comment}
\usepackage{url}
\usepackage{appendix}

\begin{document}
\title{The Digitalization of Bioassays in the \\ Open Research Knowledge Graph}
%\thanks{Supported by TIB Leibniz Information Centre for Science and Technology, the EU H2020 ERC project ScienceGRaph (GA ID: 819536)}
\author{Jennifer D'Souza\inst{1}\orcidID{0000-0002-6616-9509} \and
Anita Monteverdi\inst{2} \and
Muhammad Haris\inst{3}\orcidID{0000-0002-5071-1658} \and
Marco Anteghini\inst{4}\orcidID{0000-0003-2794-3853} \and
Kheir Eddine Farfar\inst{1}\orcidID{0000-0002-0366-4596} \and
Markus Stocker\inst{1}\orcidID{0000-0002-6616-9509} \and
Vitor A.P. Martins dos Santos\inst{4}\orcidID{0000-0002-2352-9017} \and
S\"oren Auer\inst{1,3}\orcidID{0000-0002-0698-2864}}
\authorrunning{D'Souza et al.}
\institute{TIB Leibniz Information Centre for Science and Technology, Hannover, Germany \\
\email{\{jennifer.dsouza,kheir.farfar,markus.stocker,auer\}@tib.eu}
\and
Brain Connectivity Center, IRCCS Mondino Foundation, 27100 Pavia, Italy \\
\email{anita.monteverdi01@universitadipavia.it}
\and
L3S Research Center, Leibniz University Hannover, Germany \\
\email{haris@l3s.de}
\and
Lifeglimmer GmbH, Markelstr. 38, 12163 Berlin, Germany
\email{\{anteghini,vds\}@lifeglimmer.com}
}
\maketitle              % typeset the header of the contribution
\begin{abstract}
%Biological assays have been traditionally disseminated and preserved through unstructured or semi-structured descriptions published as supplementary material to the results in Life Science scholarly articles. However, this discourse-centric representation paradigm is limited in its machine-interpretability since it does not permit automated comparison and reasoning. 
%In contrast to the traditional dissemination of bioassays as unstructured or semi-structured descriptions, this work proposes a digitalized, i.e. a semantically rich and interlinked, description of bioassays via a synergistic workflow of manual and automated methods. Specifically, the paper describes the Open Research Knowledge Graph Bioassay Semantifier (\textsc{ORKG-assays}) to semantify bioassay text as a Knowledge Graph (KG) with an AI component producing 5,514 unique property-value pairs for 103 predicates extracted from over 900 bioassays. The AI component leverages a clustering method which yielded competitive results on gold-standard data evaluations. The \textsc{ORKG-assays} module can be applied on various bioassay depositories for their unstructured data digitalization. As a result, the semantified assays can be surveyed via tabulation or chart-based visualizations of key property values in turn offering smart knowledge access to biochemists and pharmaceutical researchers in the advancement of drug development. 

\textit{\textbf{Background:}} Recent years are seeing a growing impetus in the semantification of scholarly knowledge at the fine-grained level of scientific entities in knowledge graphs. The Open Research Knowledge Graph (ORKG) \url{https://www.orkg.org/} represents an important step in this direction, with thousands of \textit{scholarly contributions} as structured, fine-grained, machine-readable data. There is a need, however, to engender change in traditional community practices of recording contributions as unstructured, non-machine-readable text. For this in turn, there is a strong need for AI tools designed for scientists that permit easy and accurate semantification of their scholarly contributions. We present one such tool, \textsc{ORKG-assays}. \textit{\textbf{Implementation:}} \textsc{ORKG-assays} is a freely available AI micro-service in ORKG written in Python designed to assist scientists obtain semantified bioassays as a set of triples. It uses an AI-based clustering algorithm which on gold-standard evaluations over 900 bioassays with 5,514 unique property-value pairs for 103 predicates shows competitive performance. \textit{\textbf{Results and Discussion:}} As a result, semantified assay collections can be surveyed on the ORKG platform via tabulation or chart-based visualizations of key property values of the chemicals and compounds offering smart knowledge access to biochemists and pharmaceutical researchers in the advancement of drug development.

\end{abstract}
\section{Introduction}

%advancement of digital libraries from access technologies to information-centered technologies
%Research for the advancement of digital libraries is widespread~\cite{dl1,dl2,dl3,dl4,dl5,dl6}. 
%Digital libraries are seen fundamentally as a resource that reconstructs the intellectual substance and services of a traditional library in digital form~\cite{seadle2007defining}. Within the scope of this definition, issues such as metadata handling, subject or keyword classification, and data storage and representation are tackled. Digital libraries, however, unlike their physical counterpart have plenty of room to keep evolving within the constraints of available information technology. In this respect, library and memory processes in research institutes, universities, and educational institutions as well as in research departments of companies and enterprises, in general, that are currently focused on document-based information and knowledge exchange will entail a transformative representational rethinking in the coming years. This is motivated, in part, by the digitalization successes seen in other information-rich publishing and communication services, such as in the general domain of encyclopedias, mail order e-commerce, map publishing, respectively. 

Mainstream publishing digital libraries such as the general domain encyclopedias, e-commerce products, maps, etc., have undergone a transformative \textit{digitalization} of their traditional document-based publication based on new means of information organization and access as Knowledge Graphs (KGs) - large semantic networks of fine-grained entities and relationships. As such the semantic knowledge is automatically customizable into knowledge views of smaller scopes given a query. This means a semantic query (e.g., using SPARQL) over a KG from semantified product descriptions, for instance, could select and aggregate similar properties such as price, manufacturer, dimensions, etc., to generate a comparison of products. Thus for digital libraries, in general, the current developments in increasing dissemination of commercial research information systems (e.g. Pure by Elsevier) and social networks (e.g. ResearchGate) as well as non-European initiatives (e.g. Open Knowledge Network \url{https://www.nitrd.gov/nitrdgroups/index.php?title=Open_Knowledge_Network} of the major US research funders) demonstrate that the transition from document-based to fine-grained, digitalized knowledge-based information flows is necessary and imminent. Historically, traditional libraries evolved to digital libraries to meet the \textit{access} needs of their patrons. In the present day, the digital library is evolving toward digitalization not just of metadata and keywords, but also of content to meet the \textit{information} needs of patrons with the promise of smart access methods to the fine-grained knowledge directly.

%focus of this work on scholarly data
%orkg focus on scholarly contributions also the focus of this work. Evolving to bioassays.
Aligned with the digitalization impetus, the Open Research Knowledge Graph (ORKG)~\cite{zenodo_orkg,orkg} digital library project addresses scholarly content digitalization as a distributed, decentralized, and collaborative structured scholarly knowledge creation process that can be powered with automated semantification modules via a continuous, ongoing development cycle of autonomously maintained AI micro-services. The focus in the ORKG is: \textit{to obtain fine-grained semantified `research results,' aka scholarly contributions, as knowledge graphs such that research progress can be made skimmable within user-definable knowledge scopes over key scientific entities and properties.} To this end, this paper supports the rapid assimiliation of digitalized knowledge in the scholarly data domain of biological assays (bioassays) by proposing an AI-based semantification service. The current coronavirus pandemic situation sheds critical light on advancing the drug development research lifecycle for which bioassays are crucial, hence we focus on this domain. A bioassay is, by definition, a standard biochemical test procedure used to determine the concentration or potency of a stimulus (physical, chemical, or biological) by its effect on living cells or tissues \cite{hoskins1962uses,irwin1953statistical}. Toward machine-interpretability, a biossay description can be represented as fine-grained semantified triples using the Bioassay ontology (BAO) \cite{bao1,bao2} with main information categories such as perturbagen, format, design, detection technology, meta target, endpoint, that need to be captured in order for them to be a meaningful semantic representation, and which imports other ontologies as well such as the Cell Line Ontology (CLO) \cite{clo}, Gene Ontology (GO) \cite{go}, and the NCBI Taxonomy (\url{https://www.ncbi.nlm.nih.gov/Taxonomy/taxonomyhome.html/}). Bioassays being highly diverse are clearly a non-trivial semantification domain posing challenges to standardizing and integrating the data with the goal to maximize their scientific and ultimately their public health impact as the assay screening results are carried forward into drug development programs. See Appendix \ref{s:bioassay} for more information.

%by comparing compounds active in one assay with those active in a second, one can make judgments about selectivity
%by downloading properties for compounds similar to a query one can investigate the behavior of a series of compounds rather than individual compounds

%There is thus a need for tools to be developed that allow easy search, access and download of information in PubChem, and in particular which allow one to move information en bloc to one's own computer for further processing. The development of PubChemSR was thus driven by the desire to have at hand such features as: • Easy search and retrieval of detailed compound, substance and bioassay information, including substructure and similarity searching • Interactive refinement of searches • Facility to export information to simple text or Microsoft Excel files and to specifically include or exclude individual data fields • The ability to easily retrieve compounds that are active or inactive (or both) in particular bioassays

\textsc{ORKG-assays}, the semantification service proposed in this paper for bioassays, will be the first Life Science domain supported by an automated semantification micro-service in the ORKG. To our knowledge, it fosters the development of the first end-to-end bioassay digitalization workflow in the overall scholarly community as well. The workflow involves four steps. \textbf{1)} Querying a bioassay depositor for their unstructured or semi-structured assays. Commonly, bioassays raw data are obtained via the PubChem depository~\cite{pubchem,pubchemsr} -- a major depositor of bioassays from various research institutes. \textbf{2)} Semantifying the assay via the \textsc{ORKG-assays} AI model. \textbf{3)} Linking the depositor provided assay cross-references to their scientific articles. And, \textbf{4)} integrating the bioassay semantic graph to the ORKG. Programmed in Python, \textsc{ORKG-assays} provides web-based and programmatic tools for semantifying bioassay texts. The semantified bioassay once entered in the ORKG is \textit{editable} via user-friendly frontend interfaces, is \textit{surveyable} via tabulations~\cite{oelen2019comparing} or 2-D chart visualizations~\cite{wiens2020towards}, and is \textit{queryable} for various scientific semantic ORKG relationships. \textsc{ORKG-assay} demonstrates high semantification performance F1 scores above 80\% and has been chosen after diverse methodological tests including the state-of-the-art, bidirectional transformer-based SciBERT model discussed in prior work~\cite{marco1,marco2,anteghini2021easy}. This module is further explained in Section \ref{s:ai-sem}. Finally, note that the ORKG will not serve as a mere mirror of other Bioassay depositories, but will itself be a unique application of a highly-structured science-wide knowledge graph of scholarly contributions which incoporates semantified bioassays as well.

Summing up, \textsc{ORKG-assays} offers a highly accurate and pragmatic semantification model alleviating unrealistic expectations on scientists to semantify their bioassays from scratch, but instead offers them a mere curatorial role of the automatic annotations. The pace with which novel bioassays are being submitted suggests that we have only begun to explore the scope of possible assay formats and technologies to interrogate complex biological systems. Thus this data domain, specifically, promises long-standing future application discovery many of which remain potentially untapped. Furthermore, inspired by the method we demonstrate, by drastically reducing the time required to semantify data for other scholarly domains as well, digitalization can be realistically advocated to become a standard part of the publication process.

\begin{figure}[!ht]
  \centering
  \includegraphics[width=\linewidth]{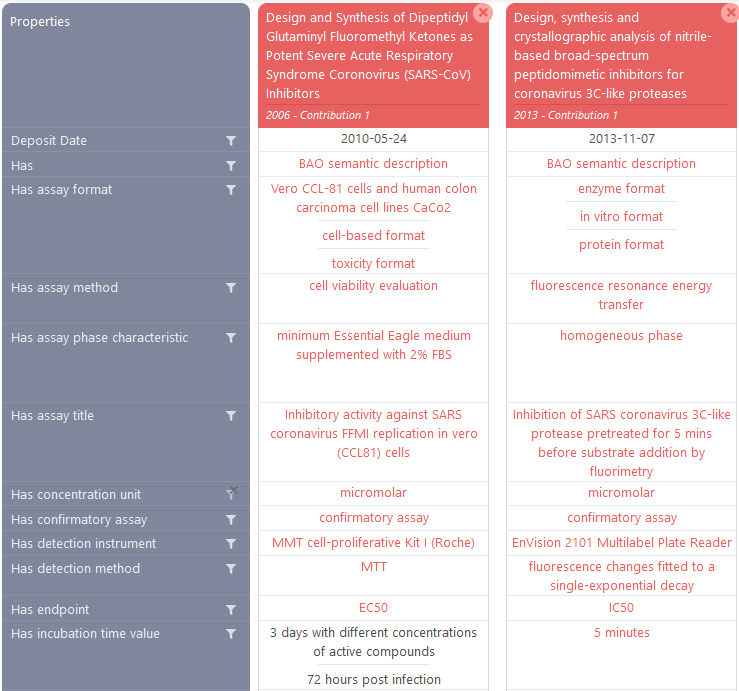}
  \caption{Two Sars-virus-related bioassays surveyed in the ORKG in terms of 12 key properties and values represented as RDF triple statements in the ORKG backend. \small{The complete survey comparing six bioassays over 21 properties can be accessed online https://www.orkg.org/orkg/comparison/R161302.}}
  \label{fig:scenario1}
\end{figure}

\section{Motivating Use Case for Digitalized Bioassays}

A biochemist wants to compare existing bioassays that have been historically performed on the SARS virus. Assuming the availability of bioassays represented as machine-readable logical triple statements, a survey over the digitalized assay KG data can, in fact, be directly computed. Concretely, see the Fig.\ref{fig:scenario1} survey of two bioassays computed over the open research knowledge graph of bioassays in the ORKG platform frontend. Note that such an information access mechanism not only directly addresses the information need of the biochemist but also alleviates their effort to otherwise have to sift through volumes of unstructured bioassay descriptions to spot the key information. Specifically, taking advantage of the ORKG, the biochemist could even dynamically compute other tailored surveys to their information need. For example, they might want to gain a comprehensive view only of those bioassays for molecules tested for the SARS virus which did not elicit a significant effect against the pathology. From the insights in such a dynamically computed view, they can consequently avoid testing the same molecule again, while focusing their attention on discovering more effective molecules. As another example, by having an expansive view on the various bioassays already tested in the literature, it should be possible to easily choose whether to repeat the experiments reproducing the same conditions of another research group or to try a new way to test compounds efficacy changing the type of bioassay or the type of cell cultures. %For this, they can compute a survey for all effective molecules against the SARS virus and compare the experimental settings. 

Taking a broader vision, \textit{experiment silos} are a problem frequently raised in research and a comparison of experimental models among research groups is not so frequent. In the context of the proposed work over bioassays, the ORKG digitalization service alleviates this problem of experimental silos by connecting all bioassays in its scholarly knowledge graph (appendix \ref{one-bioassay} illustrates the semantic description of one assay overlaid with its KG in the ORKG). Such KGs makes the bioassay data precisely \textit{findable} in accord with the FAIR standards~\cite{fair} advocated for research information. Overall, this could lead to a considerable minimization of time in research --- important especially during emergencies such as the pandemic we currently face caused by the Sars-Cov-2 virus.

\section{AI-based Bioassay Semantification}
\label{s:ai-sem}

In this section, we delve into the details of the heart of the \textsc{ORKG-assays} workflow involving the AI-based semantification module.

In general, AI-based automated scholarly KG construction, or the semantification of scholarly knowledge, is addressed by identifying and classifying entities and relations in scientific articles \cite{sys1,sys2,sys3,sys4,sys5,sys6,sys7} as sequence labeling and classification objectives, respectively. Instead, we address the problem of bioassay semantification with a clustering objective. We choose clustering from our corpus observations that bioassays with similar text descriptions are semantified with similar sets of logical statements. Thus, the bioassays could be clustered based on their text descriptions and each cluster could be collectively semantified by the labels of the trained cluster. In contrast, while entity and relation classification would also serve as sound strategies, we would unnecessarily rely on more complex and time-consuming methods. We refer the reader to our prior work~\cite{anteghini2021easy} wherein we have contrasted a classification versus a clustering objective for bioassay semantification. Last but not the least, an advantage of clustering is that it is in principle unsupervized without dependence on gold-standard data.

In the following subsections, the model formalism, implementation details, and experimental evaluation results are provided.

\subsection{Model}
\label{s:ai}

\subsubsection{Formalism.} Let $K$ be the total number of clusters of bioassays represented by the set $C = \{c_1, c_2, ..., c_K\}$. $B_{train} = \{b_1, b_2, ..., b_n\}$ corresponds to the total bioassays in the training set used to obtain optimal cluster centroids; and $V_{train} = \{v_1, v_2, ..., v_n\}$ is the vectorized representation of each bioassay to fit the clustering model. Note, $K < n$. Further, each cluster $c_x$ is associated with all the distinct logical statements of the bioassays in the respective cluster group. If cluster $c_x$ is fitted with two bioassays $b_p$ and $b_q$ in the training set, then $c_x$ is associated with a union of the logical statements of $b_{p}$ and $b_{q}$. Thus, new logical statements sets $ls_{c_x}$ associated with the $K$ clusters are formed as $\{ls_{c_1}, ls_{c_2}, ..., ls_{c_K}\}$. After the clustering model is fitted with $V_{train}$, semantification is performed. Each new bioassay $b_{test}$ is assigned based on $v_{test}$ to its closest cluster and semantified with the logical statements set of that cluster.

\subsubsection{Limitations.} \textbf{1.} \textit{The semantification of new bioassays will be limited only to the triples/statements of the training data.} Note, there is no available dataset of bioassays that completely encapsulate all labels of the BAO. Thus, if a new bioassay has new triples from the BAO, the training set would need to be expanded with additional assays annotated with the unseen triples, and the model retrained. \textbf{2.} \textit{Non-ontologized statements of bioassays cannot be captured by the clustering approach.} Other than the standardized ontologized statements, bioassays are also annotated with assay-specific statements such as ``has temperature value $\rightarrow$ 25 degree celcius'' or ``has incubation time value $\rightarrow$ 20 minute.'' Such statements require a different semantification strategy as a text reader of entities.

\subsubsection{Implementation.} Each bioassay text is first converted into vector representations. We compare two different vectorizations: 1) TF-IDF~\cite{tfidf}; and 2) SciBERT embeddings~\cite{scibert}. The TF-IDF vector is fitted on a training collection of assays. Whereas the SciBERT embeddings are directly queried for their pretrained 768 dimensional vectors. We employ the K-means clustering algorithm~\cite{kmeans}. To determine the optimal clusters size $K$, we employ the elbow optimization strategy that tries to select the smallest number of clusters accounting for the largest amount of variation in the data~\cite{syakur2018integration}.

\subsection{Experimental Setup} 

\subsubsection{Dataset.} For our model design and experimental observations, we relied on a corpus of annotated bioassays that were provided by the BAO group \cite{clark2014fast} (cloned in \url{https://github.com/jd-coderepos/bioassays-ie}). 
The dataset contains 983 human-annotated assays in all with 5,514 unique semantic statements from the BAO. On an average, each assay is annotated with 57 semantic statements with a maximum of 162 and a minimum of 7 statements. As mentioned earlier, these statements are obtained from the BAO. For in-depth information on the BAO, we refer the reader to the original papers~\cite{bao1,bao2}.

\subsubsection{Cross Validation and Metrics.} For clustering, we performed 3-fold cross validation experiments with a training/test set distribution of approximately 600 and 300 assays, respectively. The test set assays were selected such that they were unique between the folds. We measure the standard micro-precision, recall, and F1 scores for bioassay semantification per-fold experiment. The final scores are then averaged over the three folds.

\subsubsection{Evaluation Models.} \textbf{1.} \textit{Naive method.} The top-n most frequent statements are computed across the whole dataset and each bioassay is uniformly annotated with these top-n statements. \textbf{2.} \textit{Clustering.} This is the model built in this work. Additionally, we introduce a labels frequency threshold parameter within the clusters. E.g., if a threshold of 4 is applied, then the test bioassays are semantified with only the statements that appeared 4 or more times within the cluster groups when the semantic statements from the various bioassays in that cluster were aggregated. Implicitly, we see that the application of label frequency thresholds, altogether drops some statements from consideration which, at the outset, imposes a performance disadvantage for the thresholded method applications. Nevertheless, it still offers \textit{generality} versus \textit{specificity} semantification methodological effectiveness insights.  %This parameter is introduced to test for statement generality (the more frequently appearing statements within assay clusters) versus specificity (the less frequent statements within assay clusters). 

\begin{table}[!t]
\caption{Percentage bioassay semantification results by the naive method of most frequent labels assignment}
\label{results-naive}
\begin{tabular}{ccc|ccc|ccc|ccc|ccc} \hline
\multicolumn{3}{c|}{\textbf{top 10}} & \multicolumn{3}{c|}{\textbf{top 20}} & \multicolumn{3}{c|}{\textbf{top 30}} & \multicolumn{3}{c|}{\textbf{top 40}} & \multicolumn{3}{c}{\textbf{top 50}} \\
$P$ & $R$ & $F1$ & $P$ & $R$ & $F1$ & $P$ & $R$ & $F1$ & $P$ & $R$ & $F1$ & $P$ & $R$ & $F1$ \\ \hline
\multicolumn{1}{r}{58.82} & \multicolumn{1}{r}{0.02} & \multicolumn{1}{r|}{0.04} & \multicolumn{1}{r}{64.52} & \multicolumn{1}{r}{0.04} & \multicolumn{1}{r|}{0.08} & \multicolumn{1}{r}{50.85} & \multicolumn{1}{r}{0.06} & \multicolumn{1}{r|}{0.11} & \multicolumn{1}{r}{46.51} & \multicolumn{1}{r}{0.08} & \multicolumn{1}{r|}{0.15} & \multicolumn{1}{r}{43.48} & \multicolumn{1}{r}{0.09} & \multicolumn{1}{r}{0.19} \\ \hline
\end{tabular}
\end{table}

\subsection{Results and Discussion}

The empirical results are presented in Tables \ref{results-naive} and \ref{results-tfidf}, and discussed below in detail under three research questions (\textit{\textbf{RQs}}).

\textit{\textbf{RQ 1}: Is semantification by the top-n statements an effective method?} In each of the five main columns in Table \ref{results-naive}, viz. `top 10' through `top 50,' n corresponds to the number of the most frequent statements assigned for each assay. E.g., `top 10' is the 10 most common statements; `top 50' is the 50 common statements. The results show that an increase in the number of statements (`n') for semantification insignificantly increases recall but at a significant cost to precision. In light of this, we asked ourselves: \textit{could the naive method achieve greater than 50\% F1?} The answer is no. For this to occur either \textit{P} or \textit{R} has to cross the 50\% threshold while the other value be close enough to average to a 50\% \textit{F1}. But the results show that this is certainly unlikely, since the highest recall of 0.09\% (`top 50' column in Table \ref{results-naive}) is achieved at 43.48\% precision which is only steadily declining having achieved a peak value of 64.52\% at `top 20' statements. Thus, the semantification task cannot be solved by the naive method since it cannot handle the semantification pattern variations across bioassays, proved for those in our dataset.

\textit{\textbf{RQ 2}: Is clustering suitable for bioassay semantification?} Examining the bold F1 scores in Table \ref{results-tfidf} shows that it is. Note that the lowest best F1 scores among the compared parameter settings are for `Labels freq $\ge$ 4' at 0.63 and 0.66 for SciBERT and TF-IDF vectorizations, respectively. This shows the method can achieve a performance better than chance. On the other hand, the highest best F1 scores are for `Labels freq $\ge$ 1' at 0.77 and 0.83 for SciBERT and TF-IDF vectorizations, respectively, which are strong performances for practical purposes. 

\textit{\textbf{RQ 3}: What can be concluded from TF-IDF versus SciBERT vectorization?} This is a case-in-point for computing data-specific vectors. While SciBERT~\cite{scibert} is pretrained on a dataset of Computer Science and Biomedical scholarly articles, articles are still characteristically distinct from bioassay texts in terms of length and sectional organization. Bioassays are short descriptions of 1 or 2 paragraphs with either none or very few sections. Thus, we hypothesize the straightforward TF-IDF vectorization on a data source of bioassays would create better semantic representations of the data in vector space. Our hypothesis is empirically proven by the results in Table \ref{results-tfidf}, where in all experiment settings, TF-IDF vectorization outperforms the scholarly-articles-based pretrained SciBERT model. The highest F1 obtained by SciBERT is 0.77, while using TF-IDF is 0.83. Note the up and down arrows in the table reflect an increasing or decreasing scores trend. In this respect, vectorization by TF-IDF or SciBERT show similar increases/decreases.

\begin{table}[!t]
\caption{Bioassay semantification results by K-means clustering of bioassay vectorized representations}
\label{results-tfidf}
\begin{tabular}{p{1.2cm}|lll|lll|lll|lll} \hline
\multicolumn{1}{c|}{\multirow{3}{*}{\textbf{Num. of Clusters}}} & \multicolumn{6}{c|}{\textbf{Labels freq $\ge$ 4}}  & \multicolumn{6}{c}{\textbf{Labels freq $\ge$ 3}} \\
\multicolumn{1}{c|}{} & \multicolumn{3}{c}{\textsc{tf-idf}} & \multicolumn{3}{c|}{\textsc{SciBERT}} & \multicolumn{3}{c}{\textsc{tf-idf}} & \multicolumn{3}{c}{\textsc{SciBERT}} \\
\multicolumn{1}{c|}{} & $P$ & $R$ & $F1$ & $P$ & $R$ & $F1$ & $P$ & $R$ & $F1$ & $P$ & $R$ & $F1$ \\ \hline
50 & 0.48 & 0.80 & 0.60 & 0.52 & 0.70 & 0.60 & 0.40 & 0.84 & 0.54 & 0.46 & 0.77 & 0.58 \\
100 & 0.66 & 0.66 & \underline{\textbf{0.66}} $\uparrow$   & 0.69 & 0.58 & \underline{\textbf{0.63}} $\uparrow$ & 0.62 & 0.76 & 0.68 $\uparrow$ & 0.63 & 0.66 & \underline{\textbf{0.65}} $\uparrow$ \\
150 & 0.80 & 0.49 & 0.61 $\downarrow$ & 0.77 & 0.46 & 0.58 $\downarrow$ & 0.76 & 0.63 & \underline{\textbf{0.69}} $\uparrow$ & 0.73 & 0.58 & 0.64 $\downarrow$ \\
200 & 0.83 & 0.43 & 0.56 $\downarrow$ & 0.80 & 0.40 & 0.53 $\downarrow$ & 0.80 & 0.56 & 0.66 $\downarrow$ & 0.76 & 0.51 & 0.61 $\downarrow$ \\
250 & 0.86 & 0.31 & 0.45 $\downarrow$ & 0.84 & 0.33 & 0.47 $\downarrow$ & 0.85 & 0.44 & 0.58 $\downarrow$ & 0.79 & 0.43 & 0.55 $\downarrow$ \\
300 & 0.88 & 0.24 & 0.37 $\downarrow$ & 0.86 & 0.28 & 0.43 $\downarrow$ & 0.86 & 0.35 & 0.50 $\downarrow$ & 0.82 & 0.37 & 0.51 $\downarrow$ \\
350 & 0.90 & 0.15 & 0.25 $\downarrow$ & 0.90 & 0.20 & 0.32 $\downarrow$ & 0.88 & 0.27 & 0.41 $\downarrow$ & 0.87 & 0.28 & 0.42 $\downarrow$ \\
400 & 0.93 & 0.09 & 0.17 $\downarrow$ & 0.92 & 0.11 & 0.20 $\downarrow$ & 0.91 & 0.20 & 0.32 $\downarrow$ & 0.89 & 0.18 & 0.30 $\downarrow$ \\
450 & 0.94 & 0.08 & 0.14 $\downarrow$ & 0.94 & 0.08 & 0.15 $\downarrow$ & 0.93 & 0.12 & 0.22 $\downarrow$ & 0.91 & 0.12 & 0.21 $\downarrow$ \\
500 & 0.94 & 0.05 & 0.09 $\downarrow$ & 0.98 & 0.05 & 0.09 $\downarrow$ & 0.93 & 0.08 & 0.15 $\downarrow$ & 0.93 & 0.08 & 0.16 $\downarrow$ \\
550 & 0.95 & 0.03 & 0.06 $\downarrow$ & 0.99 & 0.04 & 0.07 $\downarrow$ & 0.94 & 0.04 & 0.08 $\downarrow$ & 0.97 & 0.05 & 0.09 $\downarrow$ \\
600 & 0.95 & 0.02 & 0.05 $\downarrow$ & NaN & 0.0   & NaN  & 0.96 & 0.03 & 0.06 $\downarrow$ & 0.95 & 0.01 & 0.02 $\downarrow$ \\ \hline  
\end{tabular}
\end{table}

\begin{table}[!t]
\begin{tabular}{p{1.2cm}|lll|lll|lll|lll} \hline
\multicolumn{1}{c|}{\multirow{3}{*}{\textbf{Num. of Clusters}}} & \multicolumn{6}{c|}{\textbf{Labels freq $\ge$ 2}}  & \multicolumn{6}{c}{\textbf{Labels freq $\ge$ 1}} \\
\multicolumn{1}{c|}{} & \multicolumn{3}{c}{\textsc{tf-idf}} & \multicolumn{3}{c|}{\textsc{SciBERT}} & \multicolumn{3}{c}{\textsc{tf-idf}} & \multicolumn{3}{c}{\textsc{SciBERT}} \\
\multicolumn{1}{c|}{(continued)} & $P$ & $R$ & $F1$ & $P$ & $R$ & $F1$ & $P$ & $R$ & $F1$ & $P$ & $R$ & $F1$ \\ \hline
50 & 0.32 & 0.89 &  0.47 & 0.36 & 0.83 & 0.50 & 0.19 & 0.94 & 0.31 & 0.22 & 0.90 & 0.37 \\
100 & 0.53 & 0.85 & 0.66 $\uparrow$  & 0.52 & 0.76 & 0.62 $\uparrow$ & 0.32 & 0.92 & 0.47 $\uparrow$ & 0.34 & 0.87 & 0.49 $\uparrow$ \\
150 & 0.70 & 0.79 & \underline{\textbf{0.74}} $\uparrow$  & 0.63 & 0.72 & \underline{\textbf{0.67}} $\uparrow$ & 0.54 & 0.90 & 0.68 $\uparrow$ & 0.45 & 0.85 & 0.59 $\uparrow$ \\
200 & 0.76 & 0.72 & 0.74 & 0.69 & 0.66 & 0.67 & 0.66 & 0.89 & 0.75 $\uparrow$ & 0.53 & 0.84 & 0.65 $\uparrow$ \\
250 & 0.79 & 0.65 & 0.72 $\downarrow$  & 0.73 & 0.60 & 0.66 $\downarrow$ & 0.71 & 0.88 & 0.79 $\uparrow$ & 0.59 & 0.83 & 0.69 $\uparrow$ \\
300 & 0.81 & 0.56 & 0.66 $\downarrow$ & 0.75 & 0.54 & 0.63 $\downarrow$ & 0.75 & 0.86 & 0.80 $\uparrow$ & 0.64 & 0.81 & 0.72 $\uparrow$ \\
350 & 0.84 & 0.47 & 0.60 $\downarrow$ & 0.80 & 0.44 & 0.57 $\downarrow$ & 0.78 & 0.86 & 0.82 $\uparrow$ & 0.69 & 0.80 & 0.74 $\uparrow$ \\
400 & 0.86 & 0.38 & 0.53 $\downarrow$ & 0.82 & 0.35 & 0.49 $\downarrow$ & 0.80 & 0.85 & 0.82 & 0.72 & 0.79 & 0.75 $\uparrow$ \\
450 & 0.86 & 0.27 & 0.41 $\downarrow$ & 0.84 & 0.25 & 0.39 $\downarrow$ & 0.81 & 0.85 & \underline{\textbf{0.83}} $\uparrow$ & 0.74 & 0.79 & 0.76 $\uparrow$ \\
500 & 0.88 & 0.17 & 0.28 $\downarrow$ & 0.86 & 0.17 & 0.28 $\downarrow$ & 0.82 & 0.85 & 0.83 & 0.75 & 0.78 & 0.76 \\
550 & 0.89 & 0.09 & 0.17 $\downarrow$ & 0.89 & 0.10 & 0.19 $\downarrow$ & 0.82 & 0.84 & 0.83 & 0.75 & 0.78 & \underline{\textbf{0.77}} $\uparrow$ \\
600 & 0.94 & 0.04 & 0.07 $\downarrow$ & 0.87 & 0.03 & 0.07 $\downarrow$ & 0.83 & 0.84 & 0.83 & 0.77 & 0.78 & 0.77 \\ \hline    
\end{tabular}
\end{table}

\begin{figure}[tb]
  \includegraphics[width=\textwidth]{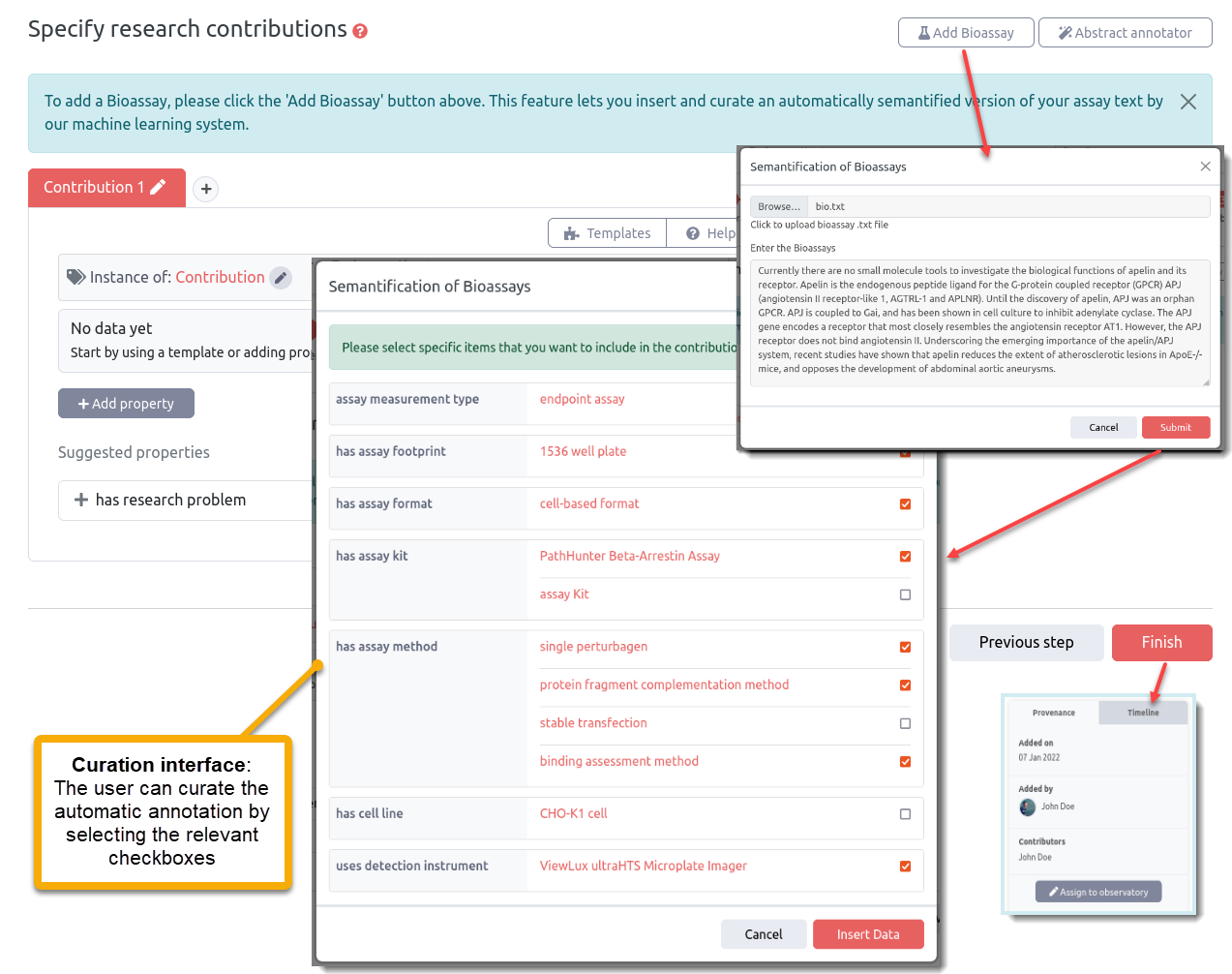}
  \caption{ORKG frontend screens for user curation of an automatically semantified bioassay.}
  \label{fig:orkg-frontend-screen}
\end{figure}

\section{Bioassay Digitalization in the ORKG}

\textsc{ORKG-assays} will now be discussed as its implementation w.r.t. the KG Lifecycle requirements \cite{kgl} consisting of the graph creation, hosting, curation, and deployment modules. The \textsc{ORKG-assays} micro-service belongs in an early stage of graph creation, i.e. when generating the graph itself. Thus, while the graph creation module handling the normalization of variously formatted graph data is beyond the scope of \textsc{ORKG-assays}, it addresses extracting the assay texts from heterogeneous bioassay depositories each with different file formats, generating a BAO-based structured graph. Such relevant details are discussed below.

%\autoref{fig:orkg-bioassay-digi-workflow} depicts the implementation of this semantification workflow over bioassays in the ORKG. Next we discuss the components in terms of the three-step approach presented earlier in Section~\ref{s:conceptual-model}.

\begin{figure}[tb]
  \includegraphics[width=0.8\textwidth]{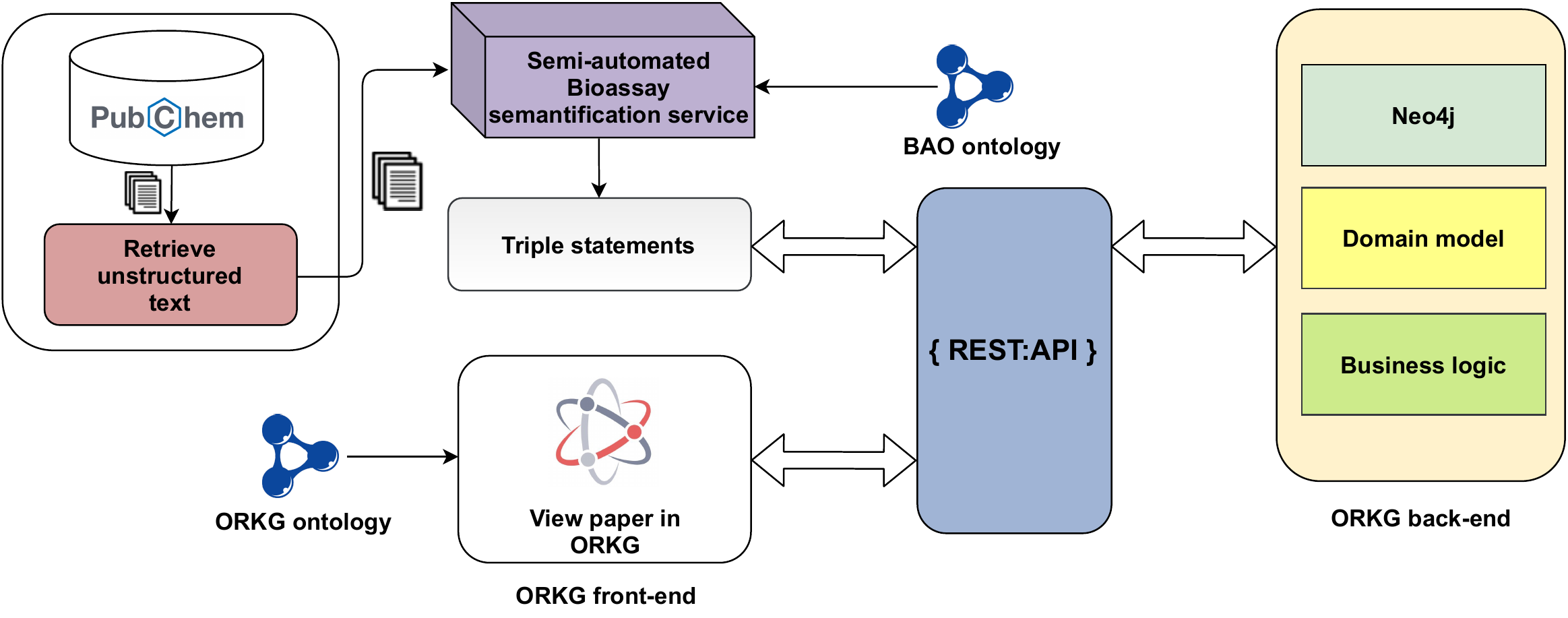}
  \caption{End-to-end \textsc{ORKG-assays} semantification pipeline which practically realizes the digitalization of digitized data shown in Fig.\ref{concept-digi-model} as a conceptual model involving data sources, data retrieval, an annotation service, and resulting triple statements.}
  \label{fig:orkg-bioassay-digi-workflow}
\end{figure}

\begin{figure}[tb]
  \includegraphics[width=\textwidth]{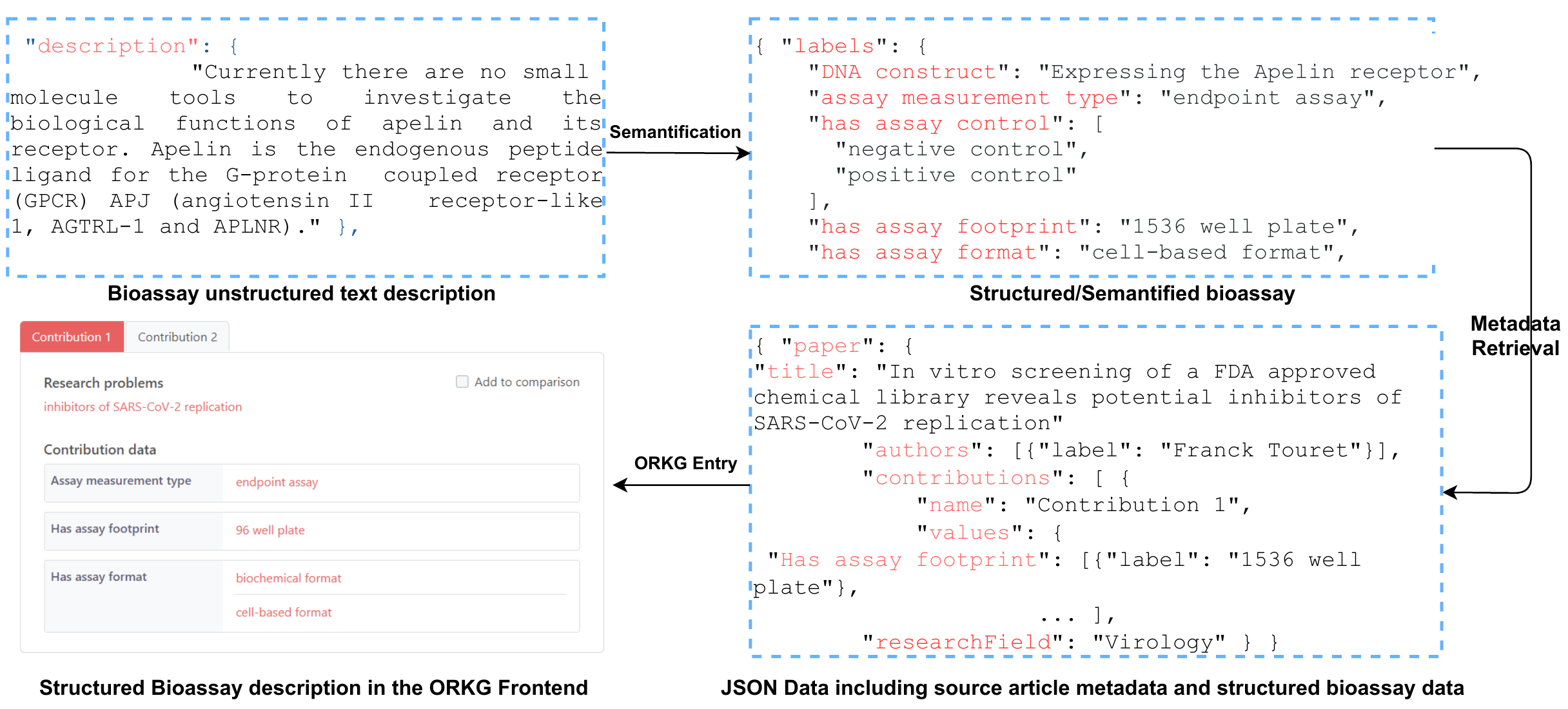}
  \caption{Conversion of an unstructured Bioassay to its equivalent structured/semantified/digitalized representation and finally presented in the ORKG frontend.}
  \label{fig:orkg-detail-bioassay-digi}
\end{figure}

\subsection{ORKG Background - Technical Details}

The design of a digital library of the future handling digitalized data should, based a common understanding of data and information between various stakeholders, integrate these technologies in the infrastructure and processes of search and knowledge exchange. This vision of knowledge-based information flows in scholarly communication requires comprehensive long-term technological infrastructure development and accompanying research. The ORKG targeting the big data space of scholarly contributions satisfies this core design objective. Foremost, as open source software (\url{https://gitlab.com/TIBHannover/orkg}), it enables a large number of partners, users and stakeholders to contribute. Programmed in the Kotlin language and Spring framework, it has cloud native design to afford scalability and extensibility. Technically, at its core, it consists of a scalable data management infrastructure with a flexible graph-based data model accessed via lightweight APIs. As exportable data formats, the service currently implements the long-established open standards RDF/RDF-Schema in accordance with FAIR Data Principles to provide maximum interoperability (\url{https://www.orkg.org/orkg/export-data}). All data and information stored in the ORKG is made available under an open license as open data and open knowledge, so that the community can use this data for integration with other services, new applications or domain-specific analyses. 
Overall, the platform consists of three main subsystems: frontend, backend, and clients. The frontend is a single-page application (SPA) providing a user-visible interface by which users can contribute, explore, and curate research data. It uses the React framework. For styling, Bootstrap with the package Reactstrap. With this technology, various flexible UI elements are currently supported such as a statement browser for curating/viewing structured scholarly contributions information (see Figs. \ref{fig:scenario1} and \ref{fig:scenario2}). It also supports graph views of the data thus offering an alternative and complementary way to interact with ORKG content (see the overlaid graph in Fig.\ref{fig:scenario2}). Next, and crucially, the ORKG supports the possibility of creating templates that specify the structure of content types, and using those templates when describing research contributions. E.g., the \url{https://www.orkg.org/orkg/template/R70247} specification for bioassays. The backend consists of several (micro-)services: the REST API, as well as components responsible for similarity, comparison, annotation, curation, and AI information extraction. These services are coordinated by a HTTP reverse proxy (Nginx). Furthermore, the REST API is built using the Hexagonal Architecture approach to facilitate splitting out backend functionality into additional micro-services easily, based on need, thus supporting easy extensibility. Moreover, this architecture permits use of different storage technologies based on suitability to the micro-service use cases as, in it, the domain logic is isolated from storage concerns. As storage systems, it currently leverages the power of property graphs (Neo4j) and relational databases (PostgreSQL). A central aspect of data storage in the ORKG is the preservation of provenance and evolution (similar to wikis), so that changes can be tracked transparently at any time. Finally, the third subsystem are wrappers around the ORKG REST API to allow direct interaction from other projects, e.g. the ORKG Python client leveraged in Jupyter notebooks.

\subsection{\textsc{ORKG-assays} Semantification Pipeline}

\subsubsection{Data preparation} This step relies on public access availability to an assay depository's querying mechanism. PubChem, reported to have over 1 million assays \cite{pubchemassociations}, is queryable via its public REST API for its bioassays where some assays have depositor-provided cross-references to scientific articles in PubMed. Depending on the depositor, the data could be returned in JSON, XML, or CSV. We implemented a specific pipeline for ``The Scripps Research Molecular Screening Center'' which returned JSON query responses. It reported nearly 1,600 bioassays. However, to prepare the data, the bioassay description-specific sections had to be located in its JSON response file and the text then extracted. The text was merged from two separate parts, viz. assay overview and assay protocol summary. We noted that this parsing heuristic can be applied to most depositor responses, although there maybe some exceptions. From the 1,600 assays, 182 contained no parseable text descriptions.

\subsubsection{Semantification} We designed this component using automated techniques and a user interface to help scientists curate their data with minimum effort. The hybrid design was based on the premise that pure machine learning is insufficiently accurate, and that expecting scientists to find the time to semantify their assays manually is unrealistic. Further, having scientists in-the-loop could help address annotations outside the scope of the training data which can later be fed back to improve the model. Note for the assays outside the scope of the training data, the semantifier returns no annotations. For the queried research institute source, of the total assays with text, 496 were semantifiable by our tool, whereas the rest were not. Charcteristically, an assay can belong to more than one paper and a paper can contain more than one assay.

\subsubsection{Building the Knowledge Graph} We leverage the ORKG to convert our structured annotations to a KG. The assay's article's PubMED metadata is first fetched, following which the digitalized bioassay is added in the form of research contributions of the paper via the ORKG KG building functions.

\noindent{\textit{Data Workflows.}} \textbf{1. Add Paper Wizard.} In the ORKG Frontend, as shown in Fig. \ref{fig:orkg-frontend-screen}, the user can add an assay by clicking the `Add Bioassay' button. The assay gets automatically semantified with the result on a screen with checkboxes enabling accept or reject user interactions. On clicking `Insert Data,' all selected statements and the user provenance form the ORKG. \textbf{2. Bulk Import via REST API.} To ingest the data in bulk, iterative calls to the ORKG REST API with article metadata and structured bioassay as contributions encapsulated in a JSON object can be made. This process is depicted in Figs.~\ref{fig:orkg-bioassay-digi-workflow} and \ref{fig:orkg-detail-bioassay-digi}.

\section{Conclusion} 

We presented \textsc{ORKG-assays} --- an end-to-end digitalization workflow of unstructured scholarly descriptions specifically addressing the problem of the digitalization of bioassays within a next-generation digital library, the ORKG. By nature of the design of the ORKG as a research infrastructure, it supports integrating the explicit semantic representation of contributions from publications with a large number of other information sources and infrastructures. They include: Metadata through services like Crossref, ORCID, etc.; Multimedia content, e.g. lectures and demos via TIB AV portal; Collaborative authoring via dokie.li etc.; Research data management, e.g. potentially the EOSC ELIXIR data platform; Open courseware, e.g. SlideWiki.org; Thesauri and Ontologies, via the OBO Foundry services API, or the NCBI, Medline, MESH taxonomies; Data linking via DataCite. Owing to this vastly interconnected graph, the resulting data from the \textsc{ORKG-assays} microservice in the ORKG will be \textit{Findable}, \textit{Accessible}, \textit{Interoperable}, and \textit{Reusable}, in other words will conform to the FAIR principles which were offered as guidelines for the creation of scholarly data~\cite{fair}.

\bibliographystyle{splncs04}
\bibliography{samplepaper}

\appendix

\section{Bioassays}
\label{s:bioassay}

A typical bioassay involves a stimulus (e.g. chemicals) applied to a subject (e.g. animals, tissues, plants). The corresponding response (e.g. death) of the subject is thereby triggered and measured. Thus, a bioassay is a type of an experiment with a domain-specific scope and purpose. Many early examples of bioassays used animals to test the carcinogenicity of chemicals. Animal bioassays have been used in all to evaluate the safety of chemicals in foods, drugs, and cosmetics; to re-create known human diseases. Aside from which, environmental bioassays are also performed where chemicals primarily associated with occupational or environmental exposures are evaluated \cite{hist-bioassay}. 

The bioassay is a centerpiece in human health risk assessment and the regulation of chemicals~\cite{hist-cancerbioassay} and are pivotal in pre-clinical research for drug development. By revealing whether a compound or biologic has the desired effect on a biological target, bioassays can drive decision-making throughout the drug discovery process, to ultimately bring new drugs to patients. Results from bioassays influence decisions about whether the chemical warrants further study in human clinical trials, potentially leading to its release onto the market. Thus, as an important precursor step, bioassays should be carefully planned to ensure they are optimized for their specific purpose. There are well-established non-commercial centers such as the Molecular Libraries Probe Production Centers Network (MLPCN)~\cite{mlpcn} or EU Openscreen \url{https://www.eu-openscreen.eu/} and commercial facilities \url{https://www.aureliabio.com/} that develop and perform bioassays. In the digital space, PubChem \url{https://pubchem.ncbi.nlm.nih.gov/} is a major depository for unstructured or semi-structured bioassay descriptions and chemical compounds. 
%If present, the PubChem data is linked to their source research in the PubMed\footnote{\url{https://pubmed.ncbi.nlm.nih.gov/}} digital library of Life Sciences scholarly articles. 
In this vein, complementing the digital space technologies, with the ORKG, we take a step further by handling \textit{digitalized} scholarly information and, in the context of this paper, that of bioassays data via the \textsc{ORKG-assays} workflow. We propose advanced information access technology based on knowledge graphs for generating a FAIR-compliant~\cite{fair} semantic description of unstructured bioassays so as to automatically compute surveys or search over their key information. A following natural question is: \textit{Is there a standard vocabulary for the semantification of unstructured bioassays?} Broadly speaking, ontologies have traditionally been used in biology to organize information within a domain and, to a lesser extent, to annotate experimental data. Consider the Gene Ontology~\cite{go}, the hundreds of ontologies in the Open Biological and Biomedical Ontologies (OBO) Foundry~\cite{obo}, as well as the National Center for Biomedical Ontologies \url{https://bioportal.bioontology.org/}. Keeping precedent, bioassay properties and values have also been comprehensively organized with the Bioassay Ontology~\cite{bao1,bao2} with information regarding assay formats (e.g. cell-based vs. biochemical), readout technologies, reagents employed, and details of the biological system interrogated. The BAO, is thus indeed leveraged as the reference vocabulary for bioassay semantification.

\newpage

\section{A Bioassay in the Open Research Knowledge Graph}
\label{one-bioassay}

\begin{figure}[!ht]
  \centering
  \includegraphics[width=\linewidth]{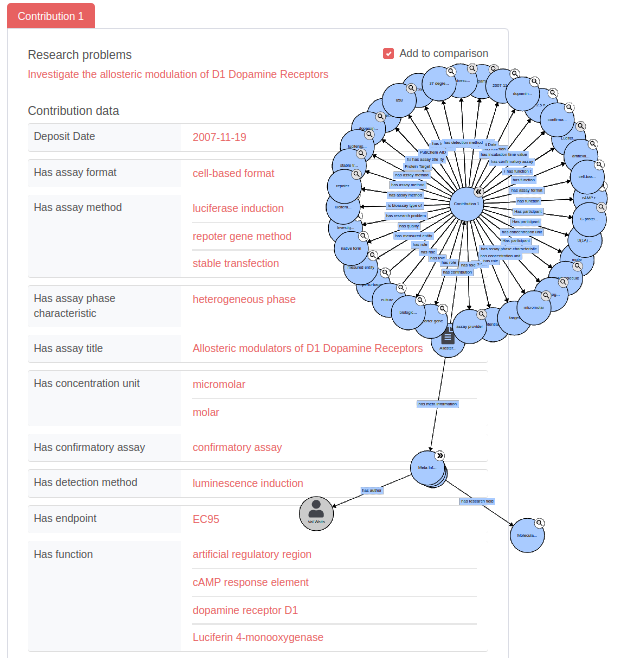}
  \caption{A semantic representation of a bioassay with an overlayed graph view of the triple statements in the ORKG frontend. Accessible online at https://www.orkg.org/orkg/paper/R48146.}
  \label{fig:scenario2}
\end{figure}

\end{document}